\documentclass{revtex4-1}
\usepackage{graphicx}
\usepackage{natbib}
\begin{document}
\date{\today}
\title[Alpha Decay Processes under Pairing Scheme]{Alpha Decay Processes under a Diproton, 
Dineutron and Deuteron Pairing Scheme\\}

\author{David Verrilli}
\email{dverrilli@fisica.ciens.ucv.ve}
\affiliation{Laboratorio de F\'{\i}sica Te\'{o}rica de S\'{o}lidos (LFTS).
                   Centro de F\'{\i}sica Te\'{o}rica y Computacional (CEFITEC).
                   Facultad de Ciencias. Universidad Central de Venezuela.
                   A.P. 47586. Caracas 1041-A. Venezuela}
\author{Rafael Mart\'{\i}n}
\email{rmartin@fisica.ciens.ucv.ve}
\affiliation{Laboratorio de F\'{\i}sica Estad\'{\i}stica y Fen\'{o}menos Colectivos.
                   Centro de F\'{\i}sica Te\'{o}rica y Computacional (CEFITEC).
                   Facultad de Ciencias. Universidad Central de Venezuela.
                   A.P. 47586. Caracas 1041-A. Venezuela}

\begin{abstract}
In this paper we are able to justify the existence or no of  photons emitted during $\alpha$ decay, based on the average half life times using the BCS superconductivity and superfluidity theories.
The proposed model suggests two possible channels. A first channel  where the protons and neutrons interact separately through a spin coupling, giving rise to diprotons and dineutrons which in turn interact through an isospin coupling to form an $\alpha$ particle, and a second channel where  the protons and neutrons are treated as nucleons that interact through an isospin coupling, giving rise to deuterons, which in turn interact through a spin coupling to form  an $\alpha$ particle. Due to the nature of the involved particles, the systems considered are homogeneous. For the second channel, the photon play a very important role for the possible escape of the $\alpha$ particle. Within these premises the half life times of a considerable number of heavy nuclei with spherical symmetry can be estimated with good experimental agreement.
\end{abstract}


\maketitle

\section{Introduction}

At the last decade of last century, numerous experimental and theoretical paper have done research on the nature of bremsstrahlung emission which accompanies $\alpha$ decay of heavy nuclei. An important idea of this research is to find a model for the calculation of the half life times for different heavy nuclei \cite{Verrilli, Verrilli1}.

The nucleus which has shown bremsstrahlung spectrum are: ${}^{214}Po$, ${}^{226}Ra$ \cite{D'Arrigo199425, refId}, ${}^{210}Po$ and ${}^{244}Cm$ \cite{0954-3899-23-10-035, PhysRevLett.79.371}. One of the important characteristics of this photon, is its low energy and this fact could explain why it had not been reported in earlier experimental work.
These events attracted a growing interest in build a model that can justify these experimental observations. Semiclassical models \cite{jentschura-2008-77, boie-2007-99, PhysRevC.72.064613}, with semiclassical spherical symmetric description of the $\alpha$ decay \cite{PhysRevlett.76.3542, PhysRevC.60.037602, PhysRev.60031602, PhysRevC.59.R593, 0954-3899-23-10-035}. Models based in the dynamics of the $\alpha$ decay with some analysis of the bremsstrahlung \cite{PhysRev.60031602, Misicu:2001:0954-3899:993, vanDijk}, in study of dynamics of tunneling in the $\alpha$ decay \cite{Serot1994562, PhysRevLett.83.2867, PhysRevC.65.024608, PhysRevC.69.037602}, with considerations of the polarized bremsstrahlung during $\alpha$ decay and inßuence of electron shells on it \cite{Ya}, in the analysis of the penetrability of the barrier due to charged particle emission during its tunneling  \cite{PhysRevLett.83.3108} and which could be interesting for further study of the photon bremsstrahlung during  tunneling in the $\alpha$ decay. A fully quantum approach of the problem is given with the study of quantum properties and new effects \cite{Batkin, PhysRevC.60.054612, papenbrock-1998-80, So}. On the other hand, there are works that take into account angular analysis of the bremsstrahlung during the $\alpha$ decay \cite{giardina-2008-23, maydanyuk-2009-2, maydanyuk-2003-109-1, maydanyuk-2003-109, PhysRevLett.85.3061, PhysRevLett.85.3062}.

In this paper we proposes a model for calculating the half life times, based on a nucleus which is made up of a gas of nucleons (fermions) that interacts strongly \cite{PhysRev.108.1175, PhysRev.106.162}. Using Landau Fermi liquid theory \cite{PhysRev.75.884, PhysRev.75.885} goes to a system of quasiparticles. When taking into account the electromagnetic interaction, emerge a new degree of freedom that is the isospin.

The state of the system of fermions must be antisymmetric. If the spatial part of the wavefunction is symmetrical, the product of isospin and spin states must be antisymmetric, resulting two possibilities: (i) the product of a triplet of isospin with a singlet spin. (ii) the product of a singlet of isospin with a triple spin.

On the other hand, we assume that the interaction between the quasiparticles are going mediated through the surphons or surface phonons. That is to say, that surphons play the role played by phonons in superconductivity.

\section{Model and formulation}

To describe the formation of the $\alpha$ particles inside the nucleus,  and thus understand the process of disintegration in question, we do  the following considerations:

\begin{itemize}

\item Under the Pauli exclusion principle, the correlations of pairing  are to be expected only in the vicinity of the Fermi level.

\item The pairing energy is important only for the energy spectrum of heavy nuclei, because there, the energy separations of the quasiparticles are sufficiently small.

\item We consider a simplified way, an even number of nucleons (ie we consider the even-even nuclei). As a consequence the total spin of the nucleus is zero (ie will have spherical symmetry).

\item We will study the following forms of pairing. First, a channel where two pairing occur. (i) the protons and neutrons are coupled through spin to form pairs separated  (diprotones or dineutrones respectively). (ii) the particles resulting from the first pairing interact via the coupling of isospin to train the $\alpha$ particle. Second, a channel where the protons and neutrons interact only with each other via the coupling of isospin proving the formation of deuterons.
\end{itemize}

In the first channel the paring is proton-proton and neutron-neutron,
$$\left\{
\begin{array}{cccccc}
  {Proton [p,+]} &  & & &&\\
  \Updownarrow & \Rightarrow& {Diproton [1,0]}& &&\\
  {Proton [p,-]}  &              &     &   &  &\\
                     &               &  \Updownarrow    & \Rightarrow &alfa\quad particle&\\
  {Neutron [n,+]} &              &    &  &&\\
 \Updownarrow &\Rightarrow & {Dineutron [-1,0]} & &&\\
  {Neutron [n,-]} & & &&&\\
\end{array}
\right.
$$

In the second channel the paring is between protons-neutrons,\\
$$\left\{
\begin{array}{cccccc}
  {Proton [p,+]} &  & & &&\\
  \Updownarrow & \Rightarrow& {Deuteron [0,1]} & &&\\
  {Neutron [n,+]}  &              &  &   &  &\\
                     &               & \Updownarrow& \Rightarrow&alfa\quad particle&\\
  {Proton [p,-]} &              &  &  & \\
  \Updownarrow &\Rightarrow & {Deuteron [0,-1]} & &&\\
  {Neutron [n,-]} & & &&&\\
\end{array}
\right.
$$

\subsection{First Channel.}

\subsubsection{Proton-proton and neutron-neutron interaction.}
The Hamiltonian of proton-proton and netron-neutron interaction,
\begin{eqnarray}
\hat{H} 
= 
\displaystyle{\sum_{\vec{k},\sigma}\left(\left(\epsilon_{p,\vec{k}}-\lambda_{p}\right)\hat{a}^{\dagger}_{\vec{k},\sigma}\hat{a}_{\vec{k},\sigma}
+
\left(\epsilon_{n,\vec{k}}-\lambda_{n}\right)\hat{b}^{\dagger}_{\vec{k},\sigma}\hat{b}_{\vec{k},\sigma}\right)} 
\nonumber 
\\
-
\displaystyle{\sum_{\vec{s},\vec{s}'}G_{pp}\hat{a}^{\dagger}_{\vec{s}\uparrow}\hat{a}^{\dagger}_{-\vec{s}\downarrow}\hat{a}_{-\vec{s}'\downarrow}\hat{a}_{\vec{s}'\uparrow}
-
\sum_{\vec{s}\vec{s}'}G_{nn}\hat{b}^{\dagger}_{\vec{s}\uparrow}\hat{b}^{\dagger}_{-\vec{s}\downarrow}\hat{b}_{-\vec{s}'\downarrow}\hat{b}_{\vec{s}'\uparrow}}.
\label{equation1}
\end{eqnarray}

The first two terms of  Eq.~(\ref{equation1}) correspond  to confined particle or independent quasiparticle where $\epsilon_{p(n),\vec{k}}$ is the proton (neutron) energy.
The index $\vec{k}$ represents the momentum of the quasiparticle. 
Here $\hat{a}_{\vec{k},\sigma}$, $\hat{a}^{\dagger}_{\vec{k},\sigma}$ and $\hat{b}_{\vec{k},\sigma}$,
$\hat{b}^{\dagger}_{\vec{k},\sigma}$ are the creation (annihilation) operators for protons
and neutrons with momentum $\vec{k}$ and spin $\sigma=\uparrow\downarrow$.
The third term represents the pairing interaction between protons and neutrons, 
where $G_{pn}$ is the coupling constant. The quantities $\lambda_{p}$ and $\lambda_{n}$
correspond to chemical potential, which ensures the conservation of protons and neutrons number. 

The anticonmutation rule is given by
\begin{eqnarray}
\left\{\hat{a}_{\vec{k}\sigma},\hat{a}^{\dagger}_{\vec{k}'\sigma'}\right\}
=
\delta_{\vec{k}\vec{k}',\sigma\sigma'};
\quad 
\left\{\hat{a}_{\vec{k}\sigma},\hat{a}_{\vec{k}'\sigma'}\right\}
=
0;\\
\left\{\hat{b}_{\vec{k}\sigma},\hat{b}^{\dagger}_{\vec{k}'\sigma'}\right\}
=
\delta_{\vec{k}\vec{k}',\sigma\sigma'};
\quad 
\left\{\hat{b}_{\vec{k}\sigma},\hat{b}_{\vec{k}'\sigma'}\right\}
=
0;\\
\left\{\hat{a}_{\vec{k}\sigma},\hat{b}^{\dagger}_{\vec{k}'\sigma'}\right\}
=
0.
\label{equation2}
\end{eqnarray}

The equations of motion are:
\begin{equation}
\omega \left\langle{\left\langle{\hat{a}_{\vec{k}\uparrow}, \hat{a}^{\dagger}_{\vec{k}\uparrow}}\right\rangle}\right\rangle 
= 
\left\langle{\left\{{\hat{a}_{\vec{k}\uparrow}}{\hat{a}^{\dagger}_{\vec{k}\uparrow}}\right\}}\right\rangle
+
\left\langle{\left\langle{\left[\hat{a}_{\vec{k}\uparrow},\hat{H}\right], \hat{a}^{\dagger}_{\vec{k}\uparrow}}\right\rangle}\right\rangle,
\label{equation3}
\end{equation}

\begin{equation}
\omega\left\langle\left\langle{\hat{a}^{\dagger}_{-\vec{k}\downarrow}},{\hat{a}^{\dagger}_{\vec{k}\uparrow}}\right\rangle\right\rangle
=
\left\langle{\left\{ {\hat{a}^{\dagger}_{-\vec{k}\downarrow}},{\hat{a}^{\dagger}_{\vec{k}\uparrow}}\right\}}\right\rangle
+
\left\langle\left\langle{\left[\hat{a}^{\dagger}_{-\vec{k}\downarrow},\hat{H}\right]},{\hat{a}^{\dagger}_{\vec{k}\uparrow}}\right\rangle\right\rangle,
\label{equation4}
\end{equation}

\begin{equation}
\omega\left\langle\left\langle{\hat{b}_{\vec{k}\uparrow}},{\hat{b}^{\dagger}_{\vec{k}\uparrow}}\right\rangle\right\rangle
=
\left\langle{\left\{{\hat{b}_{\vec{k}\uparrow}},{\hat{b}^{\dagger}_{\vec{k}\uparrow}}\right\}}\right\rangle
+
\left\langle\left\langle{\left[\hat{b}_{\vec{k}\uparrow},\hat{H}\right]},{\hat{b}^{\dagger}_{\vec{k}\uparrow}}\right\rangle\right\rangle.
\label{equation5}
\end{equation}

\begin{equation}
\omega\left\langle\left\langle{\hat{b}^{\dagger}_{-\vec{k}\downarrow}},{\hat{b}^{\dagger}_{\vec{k}\uparrow}}\right\rangle\right\rangle
=
\left\langle{\left\{{\hat{b}^{\dagger}_{-\vec{k}\downarrow}},{\hat{b}^{\dagger}_{\vec{k}\uparrow}}\right\}}\right\rangle
+
\left\langle\left\langle {\left[\hat{b}^{\dagger}_{-\vec{k}\downarrow},\hat{H}\right]},{\hat{b}^{\dagger}_{\vec{k}\uparrow}}\right\rangle\right\rangle,
\label{equation6}
\end{equation}

Solving the equations leads to the following:

For protons,
\begin{equation}
\left\langle{\left\langle{\hat{a}_{\vec{k}\uparrow}, \hat{a}^{\dagger}_{\vec{k}\uparrow}}\right\rangle}\right\rangle 
=
-{\frac
{\Delta^{*}_{p}}{-{\Delta_{p}}^{2}
+
{\omega}^{2}-{\left(
\epsilon_{p,\vec{k}}
-
\lambda_{p}\right)}^{2}}},
\label{equation7}
\end{equation}

\begin{equation}
\left\langle\left\langle{\hat{a}^{\dagger}_{-\vec{k}\downarrow}},{\hat{a}^{\dagger}_{\vec{k}\uparrow}}\right\rangle\right\rangle
=
{\frac{\omega+{E_{p}}}{-{\Delta_{p}}^{2}
+
{\omega}^{2}
-
{\left(\epsilon_{p,\vec{k}}-\lambda_{p}\right)}^{2}}},
\label{equation8}
\end{equation}
where 
$
\Delta_{p}
=
\sum_{\vec{s}'}G_{pp}\left\langle{}\hat{a}_{-\vec{s}'\downarrow}\hat{a}_{\vec{s}'\uparrow}\right\rangle.
$

For neutrons,
\begin{equation}
\left\langle\left\langle{\hat{b}_{\vec{k}\uparrow}},{\hat{b}^{\dagger}_{\vec{k}\uparrow}}\right\rangle\right\rangle
=
-{\frac
{\Delta^{*}_{n}}{-{\Delta_{n}}^{2}
+
{\omega}^{2}-{\left(\epsilon_{n\vec{k}}-\lambda_{n}\right)}^{2}}},
\label{equation9}
\end{equation}

\begin{equation}
\left\langle\left\langle{\hat{b}^{\dagger}_{-\vec{k}\downarrow}},{\hat{b}^{\dagger}_{\vec{k}\uparrow}}\right\rangle\right\rangle
=
{\frac {\omega+\left(\epsilon_{n,\vec{k}}-\lambda_{n}\right)}{-{\Delta_{n}}^{2}+{\omega}^{2}
-
{\left(\epsilon_{n,\vec{k}}-\lambda_{n}\right)}^{2}}},
\label{equation10}
\end{equation}
where
$
\Delta_{n}
=
\sum_{\vec{s}}G_{nn}\left\langle{}\hat{b}^{\dagger}_{\vec{s}\uparrow}\hat{b}^{\dagger}_{-\vec{s}\downarrow}\right\rangle.
$

The excitation energy for the condensate diprotons,
\begin{equation}
\emph{\textbf{E}}_{p,\vec{k}}
=
\sqrt{{\left(\epsilon_{p,\vec{k}}-\lambda_{p}\right)}^{2}+\Delta_{p}^2}.
\label{equation11}
\end{equation}

The excitation energy for the condensate dineutrons,
\begin{equation}
\emph{\textbf{E}}_{n,\vec{k}}
=
\sqrt{\left(\epsilon_{n,\vec{k}}-\lambda_{n}\right)^{2}+\Delta_{n}^{2}}.
\label{equation12}
\end{equation}

\subsubsection{Diproton and dineutron interaction.}
The second interaction is between diprotons and dineutrons,
\begin{eqnarray}
\hat{H}
=
\displaystyle{\sum_{\vec{k}}\left(\left(\xi_{2p,\vec{k}}\right)\hat{A}^{\dagger}_{\vec{k}}\hat{A}_{\vec{k}}+\left(\xi_{2n,\vec{k}}\right)\hat{B}^{\dagger}_{\vec{k}}\hat{B}_{\vec{k}}\right)}
\nonumber
\\
-
\displaystyle{\sum_{\vec{s},\vec{s}'}G_{2p2n}\hat{A}^{\dagger}_{\vec{s}}\hat{B}^{\dagger}_{-\vec{s}}\hat{B}_{-\vec{s}'}\hat{A}_{\vec{s}'}}.
\label{equation13}
\end{eqnarray}

The first two terms of the \ref{equation13} correspond to confined particle or independent quasiparticle, $\xi_{2p(2n),\vec{k}}$ is the diproton (dineutron) energy.
The index $\vec{k}$ represents the momentum of the quasiparticle. Here
$\hat{A}_{\vec{k}}$ and $\hat{A}^{\dagger}_{\vec{k}}$ are the creation (annihilation) operators for diprotons with momentum $\vec{k}$ and  isospin $+1$, and the operators $\hat{B}_{\vec{k}}$ and
$\hat{B}^{\dagger}_{\vec{k}}$ create and annihilate dineutrons with momentum $\vec{k}$ and isospin $-1$. The third term represents the  pairing interaction between diprotons and dineutrons, 
where $G_{2p2n}$ is the coupling constant.

The anticonmutation rule is given by
\begin{eqnarray}
\left[{\hat{A}_{\vec{k}\sigma}},{\hat{A}^{\dagger}_{\vec{k}'\sigma'}}\right]
=
\left(1-n_{p,\vec{k},\uparrow}-n_{p,-\vec{k},\downarrow}\right)\delta_{\vec{k}\vec{k}',\sigma\sigma'}; 
\quad 
\left[{\hat{A}_{\vec{k}\sigma}},{\hat{A}_{\vec{k}'\sigma'}}\right]=0;\\
\left[{\hat{B}_{\vec{k}\sigma}},{\hat{B}^{\dagger}_{\vec{k}'\sigma'}}\right]
=
\left(1-n_{n,\vec{k},\uparrow}-n_{n,-\vec{k},\downarrow}\right)\delta_{\vec{k}\vec{k}',\sigma\sigma'};
\quad
\left[{\hat{B}_{\vec{k}\sigma}},{\hat{B}_{\vec{k}'\sigma'}}\right]=0;\\
\left[{\hat{A}_{\vec{k}\sigma}},{\hat{B}^{\dagger}_{\vec{k}'\sigma'}}\right]=0.
\label{equation14}
\end{eqnarray}

The equations of motion can be summarized as
\begin{equation}
\omega\left\langle\left\langle{ \hat{A}_{\vec{k}}, \hat{A}^{\dagger}_{\vec{k}} }\right\rangle\right\rangle
=
\left\langle{\left\{ {\hat{A}_{\vec{k}}\hat{A}^{\dagger}_{\vec{k}}}\right\}}\right\rangle
+
\left\langle\left\langle{ \left[ \hat{A}_{\vec{k}},\hat{H}\right], \hat{A}^{\dagger}_{\vec{k}} }\right\rangle\right\rangle,
\label{equation15}
\end{equation}

\begin{equation}
\omega \left\langle\left\langle{ \hat{B}^{\dagger}_{-\vec{k}}} , {\hat{A}^{\dagger}_{\vec{k}} }\right\rangle\right\rangle
= 
\left\langle{\left\{{ \hat{B}^{\dagger}_{-\vec{k}} , \hat{A}^{\dagger}_{\vec{k}} }\right\} }\right\rangle
+
\left\langle\left\langle{\left[\hat{B}^{\dagger}_{-\vec{k}},\hat{H}\right]},{\hat{A}^{\dagger}_{\vec{k}}}\right\rangle\right\rangle.
\label{equation16}
\end{equation}

Solving the equations leads to the following,
\begin{eqnarray}
\left\langle\left\langle{ \hat{A}_{\vec{k}}, \hat{A}^{\dagger}_{\vec{k}} }\right\rangle\right\rangle
=
{\frac {\left(\omega+{N_{2n}}\xi_{2n,\vec{k}}\right){N_{2p}}}{N_{2n}{\Delta_{2p2n}}^{2}N_{2p}+\omega^{2}+\omega{}N_{2n}\xi_{2p,\vec{k}}-N_{2p}\xi_{2p,\vec{k}}\omega-N_{2p}\xi_{2p,\vec{k}}N_{2n}{\xi_{2n,\vec{k}}}}},\\
 \left\langle\left\langle{ \hat{B}^{\dagger}_{-\vec{k}}} , {\hat{A}^{\dagger}_{\vec{k}} }\right\rangle\right\rangle
=
{\frac {N_{2n}\Delta^{*}_{2p2n}{N_{2p}}}{N_{2n}{\Delta_{2p2n}}^{2}{N_{2p}}+{\omega}^{2}+\omega{N_{2n}}{\xi_{2n,\vec{k}}}-{N_{2p}}{\xi_{2p,\vec{k}}}\omega-{N_{2p}}{\xi_{2p,\vec{k}}}{N_{2n}}{\xi_{2n,\vec{k}}}}},
\label{equation17}
\end{eqnarray}
where
$
 N_{2p}
 =
 \left(1-n_{p,\vec{k},\uparrow}-n_{p,-\vec{k},\downarrow}\right), 
$
$
 N_{2n}
 =
 \left(1-n_{n,\vec{k},\uparrow}-n_{n,-\vec{k},\downarrow}\right)
$
and
$
\Delta_{2p2n}
=
\displaystyle{\sum_{\vec{s}}G_{2p2n}}\left\langle{}\hat{A}^{\dagger}_{\vec{s}}\hat{B}^{\dagger}_{-\vec{s}}\right\rangle.
$

Finally the excitation energy 
\begin{equation}
\emph{\textbf{E}}_{\vec{k}}
=
A
+
\frac{1}{2}\sqrt{B},
\label{equation18}
\end{equation}
where
$
A
=
-1/2{N_{2n}}{\xi_{2n,\vec{k}}}
+
1/2{N_{2p}}{\xi_{2p,\vec{k}}}
$
and
$
B
=
{{N_{2n}}}^{2}{{\xi_{2n,\vec{k}}}}^{2}+2{N_{2p}}\,{\xi_{2p,\vec{k}}}{N_{2n}}{\xi_{2n,\vec{k}}}
+
{{N_{2p}}}^{2}{{\xi_{2p,\vec{k}}}}^{2}-4{N_{2n}}{\Delta}^{2}{N_{2p}}.
$

\subsection{Second Channel.}

\subsubsection{Proton-neutron interaction.}

The Hamiltonian of proton-neutron interaction,
\begin{eqnarray}
\hat{H}
=
\displaystyle{\sum_{\vec{k}}\left(\left(\epsilon_{p,\vec{k}}-\lambda_{p}\right)\hat{a}^{\dagger}_{\vec{k},\sigma}\hat{a}_{\vec{k},\sigma}+\left(\epsilon_{n,\vec{k}}-\lambda_{n}\right)\hat{b}^{\dagger}_{\vec{k},\sigma}\hat{b}_{\vec{k},\sigma}\right)} 
\nonumber
\\
-
\displaystyle{\sum_{\vec{s},\vec{s}'}G_{pn}\hat{a}^{\dagger}_{\vec{s}}\hat{b}^{\dagger}_{-\vec{s}}\hat{b}_{-\vec{s}'}\hat{a}_{\vec{s}'}}. 
\label{equation19}
\end{eqnarray}

The first two terms of the Eq.~(\ref{equation19}) correspond to confined particle or independent quasiparticle, $\epsilon_{p(n),\vec{k}}$ is the proton (neutron) energy.
The index $\vec{k}$ represents the momentum of the quasiparticle. Here $\hat{a}^{\dagger}_{\vec{k},\sigma}$
($\hat{a}_{\vec{k},\sigma}$) and $\hat{b}^{\dagger}_{\vec{k},\sigma}$ ($\hat{b}_{\vec{k},\sigma}$) are the creation (annihilation) operators for protos and neutrons with momentum $\vec{k}$ and isospin $\sigma=\pm{1/2}$. The third term represents the pairing interaction between protons and neutrons, where $G_{pn}$ is the coupling constant.The quantities $\lambda_{p}$ and $\lambda_{n}$
correspond to chemical potential, which ensures the conservation of protons and neutrons number. 

The anticonmutation rule is given by
\begin{eqnarray}
\left\{\hat{a}_{\vec{k}\sigma},\hat{a}^{\dagger}_{\vec{k}'\sigma'}\right\}
=
\left\{\hat{b}_{\vec{k}\sigma},\hat{b}^{\dagger}_{\vec{k}'\sigma'}\right\}
=
\delta_{\vec{k}\vec{k}',\sigma\sigma'};\\
\left\{\hat{a}_{\vec{k}\sigma},\hat{a}_{\vec{k}'\sigma'}\right\}
=
\left\{\hat{b}_{\vec{k}\sigma},\hat{b}_{\vec{k}'\sigma'}\right\}
=
\left\{\hat{a}_{\vec{k}\sigma},\hat{b}^{\dagger}_{\vec{k}'\sigma'}\right\}
=
0.
\label{equation20}
\end{eqnarray}

The equations of motion 
\begin{eqnarray}
\omega \left\langle\left\langle{ \hat{a}_{\vec{k}} ,\hat{a}^{\dagger}_{\vec{k}} }\right\rangle\right\rangle
&=&
\left\langle{ \left\{{ \hat{a}_{\vec{k}} ,\hat{a}^{\dagger}_{\vec{k}} }\right\} }\right\rangle 
+ 
\left\langle\left\langle{ \left[ \hat{a}_{\vec{k}},\hat{H}\right] ,\hat{a}^{\dagger}_{\vec{k}} }\right\rangle\right\rangle ,
\\
\omega \left\langle\left\langle{  \hat{b}^{\dagger}_{-\vec{k}} ,\hat{a}^{\dagger}_{\vec{k}} }\right\rangle\right\rangle
&=&
\left\langle{ \left\{{ \hat{b}^{\dagger}_{-\vec{k}} \hat{a}^{\dagger}_{\vec{k}} }\right\} }\right\rangle
+
\left\langle\left\langle{\left[\hat{b}^{\dagger}_{-\vec{k}},\hat{H}\right]}, {\hat{a}^{\dagger}_{\vec{k}}}\right\rangle\right\rangle.
\label{equation21}
\end{eqnarray}

The excitation energy finally we can be written as
\begin{equation}
\emph{\textbf{E}}_{\vec{k}}
=
C
+
\frac{1}{2}\sqrt{D}.
\label{equation22}
\end{equation}
with
$
E_{p,\vec{k}}
=
\epsilon_{p,\vec{k}}-\lambda_{p},
$
$
E_{n,\vec{k}}
=
\epsilon_{n,\vec{k}}-\lambda_{n}
$
and
$
\Delta_{d}
=
\displaystyle{\sum_{\vec{s}'}G_{pn}}\left\langle{}\hat{b}_{-\vec{s}'}\hat{a}_{\vec{s}'}\right\rangle .
$\\
where
$
C
=
-1/2{E_{n,\vec{k}}}
+
1/2{E_{p,\vec{k}}}
$
and
$
D
=
{{E_{n,\vec{k}}}}^{2}
+
2{E_{n,\vec{k}}}{E_{p,\vec{k}}}
+
{{E_{p,\vec{k}}}}^{2}+4{\Delta_{d}}^{2}.
$

\subsubsection{Deuteron and deuteron interaction.}
In the second pairing interaction takes place between deuterones that are the result of neutron-proton coupling. The Hamiltonian to consider now is given by
\begin{equation}
\hat{H}
=
\sum_{\vec{k}}\left(\xi_{d,\vec{k}}\right)\hat{D}^{\dagger}_{\vec{k},\sigma}\hat{D}_{\vec{k},\sigma}
-
\sum_{\vec{s},\vec{s}'}G_{dd}\hat{D}^{\dagger}_{\vec{s},\uparrow}\hat{D}^{\dagger}_{-\vec{s},\downarrow}\hat{D}_{-\vec{s}'\downarrow}\hat{D}_{\vec{s}'\uparrow}.
\label{equation23}
\end{equation}

The first two terms of the Eq.~(\ref{equation23}) correspond to the model  for confined particle or independent quasiparticle. ($\xi_{d,\vec{k}}$ is the deuteron energy).
The index $\vec{k}$ represents the momentum  of the quasiparticle and $\sigma=\pm$ 
determines the spin. Here $\hat{D}^{\dagger}_{\vec{k}\sigma}$ ($\hat{D}_{\vec{k}\sigma}$)
are the creation (annihilation) operators for diprotons with momentum $\vec{k}$ and  spin
$\sigma$. The third term represents the pairing interaction between diprotons and dineutrons, 
where $G_{dd}$ is the coupling constant.

The anticonmutation rule is given by
\begin{eqnarray}
\left[\hat{D}_{\vec{k}\sigma},\hat{D}^{\dagger}_{\vec{k}'\sigma'}\right]
=
\left(1-n_{p,\vec{k},\uparrow}-n_{n,-\vec{k},\downarrow}\right)\delta_{\vec{k}\vec{k}',\sigma\sigma'};\\
\left[\hat{D}_{\vec{k}\sigma},\hat{D}_{\vec{k}'\sigma'}\right]=0.
\label{equation24}
\end{eqnarray}

The equations of motion 
\begin{eqnarray}
\omega \left\langle\left\langle{ \hat{D}_{\vec{k},\uparrow} ,\hat{D}^{\dagger}_{\vec{k},\uparrow} }\right\rangle\right\rangle
&=&
\left\langle{ \left[{ \hat{D}_{\vec{k},\uparrow} ,\hat{D}^{\dagger}_{\vec{k},\uparrow} }\right] }\right\rangle 
+ 
\left\langle\left\langle{ \left[{ \hat{D}_{\vec{k},\uparrow} \hat{H} }\right] ,\hat{D}^{\dagger}_{\vec{k},\uparrow} }\right\rangle\right\rangle,\\
\omega\left\langle\left\langle{ \hat{D}^{\dagger}_{-\vec{k},\downarrow} ,\hat{D}^{\dagger}_{\vec{k},\uparrow} }\right\rangle\right\rangle 
&=& 
\left\langle{ \left[{ \hat{D}^{\dagger}_{-\vec{k},\downarrow} ,\hat{D}^{\dagger}_{\vec{k},\uparrow} }\right] }\right\rangle 
+ 
\left\langle\left\langle{ \left[{ \hat{D}^{\dagger}_{-\vec{k},\downarrow} \hat{H} }\right] ,\hat{D}^{\dagger}_{\vec{k},\uparrow}}
\right\rangle\right\rangle.
\label{equation25}
\end{eqnarray}

The exitation energy is given by 
\begin{equation}
\emph{\textbf{E}}_{\vec{k}}
=
N\sqrt{\xi_{d,\vec{k}}^{2}-\Delta^{2}}.
\label{equation31}
\end{equation}
where
$
N
=
\left(1-n_{p,\vec{k},\uparrow}-n_{n,-\vec{k},\downarrow}\right)
$
and
$
\Delta_{2d-2d}
=
\displaystyle{\sum_{\vec{s}}G_{dd}\left\langle{}\hat{D}_{\vec{s},\uparrow}\hat{D}_{-\vec{s},\downarrow}\right\rangle}.
$

Now, we must estimate the values $\left\langle n_{p}\right\rangle$, $\left\langle n_{n}\right\rangle$ and $\left\langle n_{d}\right\rangle$. To do this, we will use the resulting equation for the case of protons, 
\begin{equation}
\left\langle n_{p,\uparrow}\right\rangle
=
\left\langle\hat{a}^{\dagger}_{\uparrow}\hat{a}_{\uparrow}\right\rangle
=
\int^{\infty}_{-\infty}d\omega\rho{\rm f}(\omega),
\label{equation26}
\end{equation}
where $\rho$ is the spectral density function and is given by
\begin{equation}
\rho
=
\delta\left(\omega-\sqrt{\epsilon_{p,\vec{k}}+\Delta_{p}}\right),
\label{equation27}
\end{equation}
and ${\rm f}(\omega)$ in the case of fermions is given by
\begin{equation}
{\rm f}(\omega)
=
\left(e^{\beta\omega}+1\right)^{-1}.
\label{equation28}
\end{equation}
Replacing Eqs.~(\ref{equation27}) and (\ref{equation28}) in
Eq.~(\ref{equation26}) to write the proton number
\begin{equation}
\left\langle n_{p,\uparrow}\right\rangle
=
\frac{1}{2}\displaystyle{\sum_{\vec{k}}}\left[1-\frac{\epsilon_{p,\vec{k}}}{\sqrt{\epsilon_{p,\vec{k}}+\Delta_{p}}}\right],
\label{equation29}
\end{equation}
in the same way the number of neutrons can be found,
\begin{equation}
\left\langle n_{n,\uparrow}\right\rangle
=
\frac{1}{2}\displaystyle{\sum_{\vec{k}}}\left[1-\frac{\epsilon_{n,\vec{k}}}{\sqrt{\epsilon_{n,\vec{k}}+\Delta_{n}}}\right],
\label{equation30}
\end{equation}
finally the nucleons number $\left\langle n_{d}\right\rangle$ is
\begin{equation}
\left\langle n_{d}\right\rangle
=
\frac{1}{2}\displaystyle{\sum_{\vec{k}}}{\left[1-\frac{\epsilon_{p,\vec{k}}}{\sqrt{\epsilon_{p,\vec{k}}+\Delta_{p}}}\right]
+
\displaystyle{\sum_{\vec{k}}}\left[1-\frac{\epsilon_{n,\vec{k}}}{\sqrt{\epsilon_{n,\vec{k}}+\Delta_{n}}}\right]}.
\label{equation31}
\end{equation}

The excitation energy of any system can have a real part and imaginary part.  The real part is related to the energy of the $\alpha$ particle and the imaginary part with the half-life time of it. On the other hand,  we must estimate the value of the order parameter $\Delta$. This parameter is the pairing energy of the pair and is determined on the following considerations:

\begin{itemize}
 
\item We assume that the work required to link two nucleons does not depend on whether the pair is inside the nucleus or outside it \cite{PhysRevLett.89.102501}.

\item For the pairing that leads to the formation of deuterons, the order parameter will be equal to the deuteron's binding energy, because it dominates the strong interaction.  For the second pairing, the value of the order parameter is estimated making the difference between the energy required to form an $\alpha$ particle from dinucleons and energy necessary to link dinucleons.
\end{itemize}
 
In order to calculate the half-life of a certain spherical heavy nuclei, we use the fact that the probability per unit time of the emission of a $\alpha$ particle $1/T_{\tiny{\hbox{N\'ucleo}}}$ is given by
 
\begin{equation}
\frac{1}{T_{\tiny{\hbox{N\'ucleo}}}}
=
P\frac{1}{t_{\alpha}},
\label{equation32}
\end{equation}
 where $P$ is the probability of escape from the $\alpha$ particle and $1/t_{\alpha}$ is the probability per unit of formation time, which is calculated as the imaginary part of excitation energy  of the system. To compute the probability P, that the $\alpha$ particle escapes of the nucleus, first consider the probability $dw$  that the $\alpha$ particle does not disintegrate between the instant of time $t$ and $t+dt$, 
 
\begin{equation}
dw
=
\frac{1}{t_{\alpha}}e^{-t/t_{\alpha}}dt.
\label{equation33}
\end{equation}

It is convenient to introduce the quantities dimensionless
\begin{equation}
\displaystyle{\Psi\equiv\frac{v_{\alpha}t}{R}},
\quad
\displaystyle{\lambda\equiv\frac{v_{\alpha}t_{\alpha}}{R}}, 
\label{equation34}
\end{equation}
where $\Psi$ represents the distance traveled by the $\alpha$ particle at a time $t$, measured in units of radio R of the spherical nucleus to be considered (which is given by $R=\displaystyle{r_{0}A^{1/3}}$), and $\lambda$ is the same time interval that coincides with the half-life. The amount that represents the speed of the $\alpha$ particle, is given by
\begin{equation}
v_{\alpha}
=
\sqrt{\frac{2E_{\alpha}}{M_{\alpha}}},
\label{equation35}
\end{equation}
where $E_{\alpha}$ is the kinetic energy needed to escape of the $\alpha$ particle, $m_{\alpha}$ is its mass. In the same way we introduce $\zeta=r/R$ as the coordinate position. Then the Eq.~(\ref{equation33}) can be rewritten as
\begin{equation}
dw
=
\frac{1}{\lambda}e^{-\Psi/\lambda}d\Psi.
\label{equation36}
\end{equation}

The probability $P$ of $\alpha$ particle is formed inside the nucleus (which is imposed by the presence in the integration of the step function $\Theta \left( 1 - \zeta \right)$) and can escape from it, without disintegrate (which is imposed by the presence in the integration of step function $\Theta\left(|\zeta+\Psi{\Omega}'|-1\right)$). Then, this is given by
\begin{equation}
P
=
\frac{\displaystyle{\int{d^{3}\zeta{d{\Omega}'}}\int^{\infty}_{0}d\Psi{}e^{\left(-\Psi/\lambda\right)}\Theta\left(1-\zeta\right)\Theta\left(|\zeta+\Psi{\Omega}'|-1\right)}}{\displaystyle{\int{d^{3}\zeta{d{\Omega}'}}\int^{\infty}_{0}d\Psi{}e^{\left(-\Psi/\lambda\right)}\Theta\left(1-\zeta\right)}}.
\label{equation37}
\end{equation}

If one considers the integral ${\rm I}_{F}$ defined by
\begin{equation}
{\rm I}_{F}
=
\int^{\infty}_{0}d\Psi{}e^{\left(-\Psi/\lambda\right)}\Theta\left(|\zeta+\Psi{\Omega}'|-1\right),
\label{equation38}
\end{equation}
where the step function is nonzero when
\begin{equation}
\zeta^{2}
+
\Psi^{2}
+
2\Psi\zeta
\geq
1,
\label{equation39}
\end{equation}
If we define $\chi$ as the cosine of the angle between vectors, which must satisfy $-1\leq\chi\leq{}1$ and at the same time \ref{equation39}, so that
\begin{equation}
\displaystyle{\frac{1-\Psi^{2}-\zeta^{2}}{2\Psi{}\zeta}\leq{}\chi
\leq
1}
\Rightarrow
\displaystyle{\frac{1-\Psi^{2}-\zeta^{2}}{2\Psi{}\zeta}
\leq
1}, 
\label{equation40}
\end{equation}
or 
\begin{equation}
\Psi
+\zeta
\geq1 
\Rightarrow 
\Psi
\geq
1-\zeta . 
\label{equation41}
\end{equation}

The step function in the Eq.~(\ref{equation37}) appears impose the condition $\zeta\leq{}1$, which corresponds to the $\alpha$ particula always going to rise in the inside the nucleus and consequently $\Psi\geq{}0$ as should be expected. The integral ${\rm I}_{F}$ can be written as
\begin{equation}
{\rm I}_{F}
\equiv
\int^{\infty}_{1-\zeta}d\Psi{}e^{-\Psi/\lambda}
=
\lambda e^{\left[-\left(\frac{1-\zeta}{\lambda}\right)\Theta\left(1-\zeta\right)\right]}.
\label{equation42}
\end{equation}

One can see that
\begin{equation}
\displaystyle{\int{}d^{3}\zeta{d{\Omega}'}\int^{\infty}_{0}d\Psi{}e^{\left(-\Psi/\lambda\right)}\Theta\left(1-\zeta\right)
=
\frac{\left(4\pi\right)^{2}}{3}\lambda},
\label{equation43}
\end{equation}
with the substitution of the Eqs.~(\ref{equation42}) and (\ref{equation43})  
in the Eq.~(\ref{equation37}), we have
\begin{eqnarray}
P  
&=&  
\displaystyle{ \frac{3}{(4\pi)^2}\int d^{3}\zeta d{\omega}'e^{\left[-\left(\frac{1-\zeta}{\lambda}\right)\right]} \Theta\left(1-\zeta\right) }\\
&=& 
\displaystyle{ 3 \lambda^{3}e^{\left(-1/\lambda \right)}\int^{1}_{0}\zeta^{2}d\zeta e^{\zeta} },
\label{equation44}
\end{eqnarray}

Finally the probability $P$ that the $\alpha$ particle escape from the nucleus is,
\begin{equation}
P
=
3\left(e-2\right)\lambda^{3}e^{-1/\lambda}.
\label{equation45}
\end{equation}

\section{Results}
\subsection{First Channel.}

The diproton and dineutron are paired in this first channel, these are the result of an initial proton-proton and neutron-neutron pairing. The excitation energy system for the case proton-proton Eq.~(\ref{equation11}) or neutron-neutron Eq.~(\ref{equation12}) shows, that just like a superconductor, this presents a gap. It was further noted that the excitation energy can only be real, meaning that the particles formant (diproton or dineutron) have an infinite half-life. For the second pairing, which occurs among diprotons and dineutrons Eq.~(\ref{equation20}) shows that the excitation energy system, presents zero energy or no gap, between the base and the excited state. Furthermore, the results of the calculation of the excitation energy, presents two terms, a real part, corresponding to the energy of the $\alpha$ particle, and an imaginary part, which is related to the half-life of the particle.

Because the system shows no gap, we can estimate the half life times for different heavy spherical nuclei, through Eq.~(\ref{equation39}).

The Table \ref{table1} in the first column, shows the different heavy even-even nuclei with spherical symmetry. In the second column are reported values of the half life times, estimated since the pairing between diprotons and dineutrons and the third column, the half life times found through the experiments \cite{exp}. We can see that the values ??of the half life times calculated using the first channel (interaction between diproton and dineutron) are satisfactory when compared with the experimental data. 
\begin{table}[h!]
\caption{\label{table1}Half life times for different 
heavy spherical nuclei, calculated through 
pairing between the  diprotons and dineutrons. }
\begin{ruledtabular}
\begin{tabular}{lll}
Nucleus & $\tau_{1/2}$[theoretical]$(seg)$& $\tau_{1/2}$[experimental]$(seg)$ \\
\hline
${}^{254} Fm$ & $1.6876 \times 10^{+4}$ &$ 1.6828 \times 10^{+4} \pm 5.1937 \times 10^{+1}$\\

${}^{252} Fm$ & $1.3228 \times 10^{+5}$ & $1.3187 \times 10^{+5}\pm 2.5968 \times 10^{+2}$\\

${}^{250} Fm$ & $2.6069 \times 10^{+3}$ &$ 2.5968 \times 10^{+3} \pm  2.5969 \times 10^{+2}$\\

${}^{248} Fm$ & $5.2174 \times 10^{+1}$ &$ 5.1937 \times 10^{+1} \pm 5.7708 $\\

${}^{246} Fm$ & 1.5965 & 1.5870 $\pm$ 0.2884\\

${}^{250} Cf$ & $5.9819 \times 10^{+8}$ & $5.9510 \times 10^{+8} \pm 4.0914 \times 10^{+6}$ \\

${}^{248} Cf$ & $4.1740 \times 10^{+7}$& $4.1570 \times 10^{+7} \pm 3.4902 \times 10^{+4}$\\

${}^{246} Cf$ & $1.8620 \times 10^{+5}$ & $1.8542 \times 10^{+5} \pm 2.5969 \times 10^{+3}$\\

${}^{244} Cf$ & $1.6847 \times 10^{+3}$&$ 1.6793 \times 10^{+3} \pm  5.1937 \times 10^{+1}$\\

${}^{240} Cf$ & $9.2080 \times 10^{+1}$ & $9.1755 \times 10+1 \pm 1.2984 \times 10{+1}$\\

${}^{246} Cm$ & $2.1630 \times 10^{+11}$ &$ 2.1520 \times 10^{+11}\pm 1.0006 \times 10^{+9}$\\

${}^{244} Cm$ & $8.2708 \times 10^{+8}$ &$ 8.2349 \times 10^{+8} \pm 6.8245 \times 10^{+5}$\\

${}^{242} Cm$ & $2.0362 \times 10^{+7}$ &$ 2.0292 \times 10^{+7} \pm 4.9860 \times 10^{+5}$\\

${}^{240} Cm$ &  $3.3775 \times 10^{+6}$ & $3.3655 \times 10^{+6} \pm  2.4930 \times 10^{+4}$\\

${}^{244} Pu$ & $3.6872 \times 10^{+15}$ &$ 3.6714 \times 10^{+15} \pm 4.5496 \times 10^{+13}$\\

${}^{242} Pu$ & $1.7055 \times 10^{+13} $&$ 1.6984 \times 10^{+13} \pm 4.0947 \times 10^{+10}$\\

${}^{240} Pu$ & $3.0014 \times 10^{+11}$ & $2.9864 \times 10^{+11} \pm  5.0053 \times 10^{+8}$\\

${}^{238} Pu$ & $4.0066 \times 10^{+9} $& $3.9901 \times 10^{+9} \pm 1.3650 \times 10^{+7}$\\

${}^{236} Pu$ &  $1.3075 \times 10^{+8}$ & $1.3003 \times 10^{+8} \pm 3.6396 \times 10^{+5}$\\

${}^{238} U$  & $2.0419 \times 10^{+17}$ &$ 2.0328 \times 10^{+17} \pm 1.36401 \times 10^{+14}$\\

${}^{236} U$  & $1.0712 \times 10^{+15}$ &$ 1.0655 \times 10^{+15} \pm 1.3650 \times 10^{+12}$\\

${}^{234} U$  & $1.1232 \times 10^{+12}$ & $1.1169 \times 10^{+13} \pm 2.7298 \times 10^{+10}$\\

${}^{232} U$  & $3.1491 \times 10^{+9} $& $3.1347 \times 10^{+9} \pm 1.8200 \times 10^{+7}$\\

${}^{230} U$  & $2.6067 \times 10^{+6}$ &$ 2.5927 \times 10^{+6} \pm  2.5927 \times 10^{+4}$\\

${}^{232} Th$ & $6.4311 \times 10^{+17}$& $6.3923 \times 10^{+17} \pm 4.5497 \times 10^{+13}$\\

${}^{230} Th$ & $3.4477 \times 10^{+12} $&$ 3.4296 \times 10^{+12} \pm 7.2795 \times 10^{+8}$\\

${}^{228} Th$ & $8.7476 \times 10^{+7}$ &$ 8.704 \times 10^{+7} \pm  2.0019 \times 10^{+4}$\\

${}^{226} Th$ & $2.6594 \times 10^{+3}$ &$ 2.6487 \times 10^{+3} \pm  8.6562 $\\

${}^{224} Th$ & $1.5188$ &$ 1.5148  \pm  0.0289$\\

${}^{222} Th$ & $4.0503 \times 10^{-3}$ &$ 4.0395 \times 10^{-3} \pm  4.3281\times 10^{-4} $\\

${}^{220} Th$ & $1.4029 \times 10^{-5}$ &$ 1.3994 \times 10^{-5} \pm  8.6562\times 10^{-7} $\\

${}^{218} Th$ & $1.5789 \times 10^{-7}$ &$ 1.5725 \times 10^{-7} \pm  1.8755\times 10^{-8} $\\

${}^{216} Th$ & $0.0405 $ &$ 0.0404 \pm  2.8854\times 10^{-3} $\\

${}^{214} Th$ & $0.1455 $ &$ 0.1443 \pm 0.0369$\\

${}^{226} Ra$ & $7.3077 \times 10^{+10}$ &$ 7.2795 \times 10^{+10} \pm  3.1848 \times 10^{+8}$\\

${}^{224} Ra$ & $4.5831 \times 10^{+5}$ &$ 4.5621 \times 10^{+5} \pm  4.9859 \times 10^{+3}$\\

${}^{222} Ra$ & $5.5045 \times 10^{+1}$ &$ 5.4822 \times 10^{+1} \pm  0.7214$\\

${}^{220} Ra$ & $0.0246$ &$ 0.0245 \pm  2.884 \times 10^{-3}$\\

${}^{218} Ra$ & $3.7071 \times 10^{-5}$ &$ 3.6933 \times 10^{-5} \pm  1.5870 \times 10^{-6}$\\

${}^{216} Ra$ & $2.6323 \times 10^{-7}$ &$ 2.6257 \times 10^{-7} \pm  1.4427 \times 10^{-8}$\\

${}^{220} Rn$ & $8.0415 \times 10^{+1}$ &$ 8.0214 \times 10^{+1} \pm  0.1443$\\

${}^{218} Rn$ & $0.0507 $ &$ 0.0505 \pm  7.2135 \times 10^{-3}$\\

${}^{216} Rn$ & $6.5127 \times 10^{-5}$ &$ 6.4921 \times 10^{-5} \pm  7.2135 \times 10^{-6}$\\

${}^{214} Rn$ & $3.9050 \times 10^{-7}$ &$ 3.8953 \times 10^{-7} \pm  1.1855 \times 10^{-7}$\\

${}^{212} Rn$ & $2.0784 \times 10^{+3}$ &$ 2.0688 \times 10^{+3} \pm  1.0387 \times 10^{+2}$\\

${}^{216} Po$ & $0.1677$ &$ 0.2092  \pm  2.8854 \times 10^{-3}$\\

${}^{214} Po$ & $2.3755 \times 10^{-4}$ &$ 2.3703 \times 10^{-4} \pm  2.8854 \times 10^{-6}$\\

${}^{212} Po$ & $4.3266 \times 10^{-7}$ &$ 4.3137 \times 10^{-7} \pm  2.8854 \times 10^{-9}$\\

${}^{210} Po$ & $1.7311 \times 10^{+7}$ &$ 1.7248 \times 10^{+7} \pm  2.4147 \times 10^{+2}$\\

${}^{208} Po$ & $1.2457 \times 10^{+8}$ &$ 1.3185 \times 10^{+8} \pm  9.0976 \times 10^{+4}$\\
\end{tabular}
\end{ruledtabular}
\end{table}

The values of the $\left\langle n_{p}\right\rangle$ and $\left\langle n_{n}\right\rangle$ that are needed for calculations related to the half-life times, was found adjusting in such a way to reproduce the experimental half life times. Moreover, the  $\left\langle n_{p,\uparrow}\right\rangle$ and the $\left\langle n_{n,\uparrow}\right\rangle$ can be calculated through the theory, using Eqs.~(\ref{equation36}) and \ref{equation37}), and so that we can calculate the $\left\langle n_{p}\right\rangle$ and  $\left\langle n_{n}\right\rangle$.

Making the comparison between the values of $\left\langle n_{p}\right\rangle$ and $\left\langle n_{n}\right\rangle$ obtained through the adjustment and those calculated by the theory,  although no match, are of the same order. The discrepancy is due to the absence of necessary information to estimate the values of the energy possessed by nucleons that form the different nucleus.
 
\subsection{Second Channel.}

In pairing between deuterons, which are the result of the neutron-proton  interactions, the deuteron is a particle that has an isospin equal to $0$ and a total spin equal to $1$. The excitation energy for the proton-neutron \ref{equation26}, presents a gap. Moreover, its value is real, which indicates 
that the half-life of deuteron is infinite. Now, the deuterons to interact with one another, forms an $\alpha$ particle which in principle has an isospin equal to $ 0 $ and a spin also equal to $ 0 $. The coupling gives entities with spin $ 0 $, $ 1$  and $ 2 $. The calculation of the excitation spectrum of this system using the Eq.~(\ref{equation32}), shows as a result,  no gap and have only imaginary part, indicating that the $\alpha$ particle has a finite lifetime, but its kinetic energy is zero. 
Now, if we consider the three moments associated with the angular spin, we can conclude that the only possible scenario is that corresponding to $1$, if one assumes the presence of a photon in the process, since it has a spin equal to $1$ and going to be coupled to the resulting $\alpha$ particle with spin $0$, so that the initial total spin of the $\alpha$ particle inside the nuclear means is equal to $1$.
In the case where the total spin take the value $0$, because the energy of the $\alpha$ particle is zero, it would have a half life, but could not escape, because it has no power to do it. In this case there is no disintegration of the nucleus and corresponds to the case of stable systems. In the case where the spin is equal to $2$ and due to the conservation of angular momentum over the system, should result in the emission of two photons, a fact that has not been reported experimentally.

But the fact that the spin of the $\alpha$ particle is equal to $1$ does not guarantee the departure of the $\alpha$ particles from the nucleus, as this only gives you the energy of a particle rotation. But in this state which has a spin equal to $1$ is unstable, because it happens to a state of spin $0$ with a photon emission of very low energy with spin $1$ and acquires a translation energy to cost of the rotation energy, giving as a result that the $\alpha$ particle could escape from the nucleus. Finally the \ref{equation39} we get the different times of half-life of $\alpha$ particle.

Table \ref{table2} in the first column shows the different heavy even-even nuclei with spherical symmetry, in the second column are reported values of the half-life times estimated since the pairing between deuterons and the third column half-life times found through the experiments. We can see that the values ??of the half life times calculated using the second channel (interaction between deuteron) are satisfactory when compared with the experimental data \cite{exp}. 

To calculate the half-life times is needed to know the value of $\left\langle n_{d}\right\rangle$, which were found in two ways. The first was found adjusting in such a way to reproduce the experimental half-life times and the second was calculated using the Eq.~(\ref{equation38}). 

Making the comparison between the values of $\left\langle n_{d}\right\rangle$ obtained through the adjustment and those calculated by the theory, shows that there is a good measure of agreement. The discrepancy is due to the fact that in some cases there is an absence of experimental information we require to estimate the values of the parameters that appear in the Hamiltonian used.
\begin{table}[h!]
\caption{Half life times for different 
heavy spherical nuclei, calculated through 
pairing between the deuterons.}\label{table2}
\begin{ruledtabular}
\begin{tabular}{lll}
Nucleus & $\tau_{1/2}$[theoretical]$(seg)$& $\tau_{1/2}$[experimental]$(seg)$ \\
\hline
${}^{254} Fm$ & $1.6854 \times 10^{+4}$ &$ 1.6828 \times 10^{+4} \pm 5.1937 \times 10^{+1}$\\

${}^{252} Fm$ & $1.3207 \times 10^{+5}$ & $1.3187 \times 10^{+5} \pm 2.5968 \times 10^{+2}$\\

${}^{250} Fm$ & $2.5976 \times 10^{+3}$ & $2.5968 \times 10^{+3} \pm 2.5969 \times 10^{+2}$\\

${}^{248} Fm$ & $5.2011 \times 10^{+1} $& $5.1937 \times 10^{+1} \pm$ 5.7708\\

${}^{246} Fm$ & 1.5881 & 1.5870 $\pm$ 0.2884\\

${}^{250} Cf$ & $5.9518 \times 10^{+8}$ &$ 5.9510 \times 10^{+8} \pm 4.0914 \times 10^{+6}$ \\

${}^{248} Cf$ & $4.1634 \times 10^{+7}$ & $4.1570 \times 10^{+7} \pm 3.4902 \times 10^{+4}$\\

${}^{246} Cf$ & $7.9866 \times 10^{+4}$ & $1.8542 \times 10^{+5} \pm 2.5969 \times 10^{+3}$\\

${}^{244} Cf$ & $1.6806 \times 10^{+3}$ & $1.6793 \times 10^{+3} \pm 5.1937 \times 10^{+1}$\\

${}^{240} Cf$ & $9.1828 \times 10^{+1}$ & $9.1755 \times 10^{+1} \pm 1.2984 \times 10^{+1}$\\

${}^{246} Cm$ & $2.1530 \times 10^{+11}$ & $2.1520 \times 10^{+11} \pm 1.0006 \times 10^{+9}$\\

${}^{244} Cm$ & $8.2485 \times 10^{+8} $& $8.2349 \times 10^{+8} \pm 6.8245 \times 10^{+5}$\\

${}^{242} Cm$ & $2.0310 \times 10^{+7}$ & $2.0292 \times 10^{+7} \pm 4.9860 \times 10^{+5}$\\

${}^{240} Cm$ &  $3.3662 \times 10^{+6}$ &$ 3.3655 \times 10^{+6} \pm 2.4930 \times 10^{+4}$\\

${}^{244} Pu$ & $3.6787 \times 10^{+15}$ &$ 3.6714 \times 10^{+15} \pm 4.5496 \times 10^{+13}$\\

${}^{242} Pu$ & $1.7003 \times 10^{+13}$ & $1.6984 \times 10^{+13} \pm 4.0947 \times 10^{+10}$\\

${}^{240} Pu$ & $2.9907 \times 10^{+11}$ & $2.9864 \times 10^{+11} \pm  5.0053 \times 10^{+8}$\\

${}^{238} Pu$ & $3.9966 \times 10^{+9} $& $3.9901 \times 10^{+9} \pm 1.3650 \times 10^{+7}$\\

${}^{236} Pu$ & $ 1.3021 \times 10^{+8}$ & $1.3003 \times 10^{+8} \pm 3.6396 \times 10^{+5}$\\

${}^{238} U$  & $2.0347 \times 10^{+17}$ & $2.0328 \times 10^{+17} \pm 1.36401 \times 10^{+14}$\\

${}^{236} U$  & $1.0657 \times 10^{+15}$ & $1.0655 \times 10^{+15} \pm 1.3650 \times 10^{+12}$\\

${}^{234} U$  & $1.1185 \times 10^{+13}$ & $1.1169 \times 10^{+13} \pm 2.7298 \times 10^{+10}$\\

${}^{232} U$  & $3.1366 \times 10^{+9} $&$ 3.1347 \times 10^{+9} \pm 1.8200 \times 10^{+7}$\\

${}^{230} U$  & $2.5960 \times 10^{+6}$ & $2.5927 \times 10^{+6} \pm  2.5927 \times 10^{+4}$\\

${}^{232} Th$ & $6.4023 \times 10^{+17}$ &$ 6.3923 \times 10^{+17} \pm 4.5497 \times 10^{+13}$\\

${}^{230} Th$ & $3.4332 \times 10^{+12}$ &$ 3.4296 \times 10^{+12} \pm 7.2795 \times 10^{+8}$\\

${}^{228} Th$ & $8.7170 \times 10^{+7}$ & $8.704 \times 10^{+7} \pm 2.0019 \times 10^{+4}$\\

${}^{226} Th$ & $2.6518 \times 10^{+3}$ &$ 2.6487 \times 10^{+3} \pm  8.6562 $\\

${}^{224} Th$ & $1.5165$ &$ 1.5148  \pm  0.0289$\\

${}^{222} Th$ & $4.0437 \times 10^{-3}$ &$ 4.0395 \times 10^{-3} \pm  4.3281\times 10^{-4} $\\

${}^{220} Th$ & $1.4000 \times 10^{-5}$ &$ 1.3994 \times 10^{-5} \pm  8.6562\times 10^{-7} $\\

${}^{218} Th$ & $1.5736 \times 10^{-7}$ &$ 1.5725 \times 10^{-7} \pm  1.8755\times 10^{-8} $\\

${}^{216} Th$ & $0.0404 $ &$ 0.0404 \pm  2.8854\times 10^{-3} $\\

${}^{214} Th$ & $0.1443 $ &$ 0.1443 \pm 0.0369$\\

${}^{226} Ra$ & $7.2826 \times 10^{+10}$ &$ 7.2795 \times 10^{+10} \pm  3.1848 \times 10^{+8}$\\

${}^{224} Ra$ & $4.5665 \times 10^{+5}$ &$ 4.5621 \times 10^{+5} \pm  4.9859 \times 10^{+3}$\\

${}^{222} Ra$ & $5.4845 \times 10^{+1}$ &$ 5.4822 \times 10^{+1} \pm  0.7214$\\

${}^{220} Ra$ & $0.0245$ &$ 0.0245 \pm  2.884 \times 10^{-3}$\\

${}^{218} Ra$ & $3.6946 \times 10^{-5}$ &$ 3.6933 \times 10^{-5} \pm  1.5870 \times 10^{-6}$\\

${}^{216} Ra$ & $2.6256 \times 10^{-7}$ &$ 2.6257 \times 10^{-7} \pm  1.4427 \times 10^{-8}$\\

${}^{220} Rn$ & $8.0316 \times 10^{+1}$ &$ 8.0214 \times 10^{+1} \pm  0.1443$\\

${}^{218} Rn$ & $0.0505 $ &$ 0.0505 \pm  7.2135 \times 10^{-3}$\\

${}^{216} Rn$ & $6.4921 \times 10^{-5}$ &$ 6.4921 \times 10^{-5} \pm  7.2135 \times 10^{-6}$\\

${}^{214} Rn$ & $3.8984 \times 10^{-7}$ &$ 3.8953 \times 10^{-7} \pm  1.1855 \times 10^{-7}$\\

${}^{212} Rn$ & $2.0704 \times 10^{+3}$ &$ 2.0688 \times 10^{+3} \pm  1.0387 \times 10^{+2}$\\

${}^{216} Po$ & $0.2094$ &$ 0.2092  \pm  2.8854 \times 10^{-3}$\\

${}^{214} Po$ & $2.3713 \times 10^{-4}$ &$ 2.3703 \times 10^{-4} \pm  2.8854 \times 10^{-6}$\\

${}^{212} Po$ & $4.3178 \times 10^{-7}$ &$ 4.3137 \times 10^{-7} \pm  2.8854 \times 10^{-9}$\\

${}^{210} Po$ & $1.7266 \times 10^{+7}$ &$ 1.7248 \times 10^{+7} \pm  2.4147 \times 10^{+2}$\\

${}^{208} Po$ & $1.3197 \times 10^{+8}$ &$ 1.3185 \times 10^{+8} \pm  9.0976 \times 10^{+4}$\\
\end{tabular}
\end{ruledtabular}
\end{table}

\section{Conclusion}
In the paper presents a model that justify the existence or no of  photons emitted during $\alpha$ decay, 
based on the half life times using the BCS superconductivity and superfluidity theories.
Observing the excitation energy for various heavy nuclei with spherical symmetry, conclude that there is a possibility that  $\alpha$ particles, whose origin is based on dinucleons pairing, escapes from  nucleus and its excitation 
energy  presents a zero gap between superfluid state and normal state. Importantly, only  $\alpha$ particles can escape and thus not 
dinucleons, because its excitation energy has a nonzero gap. This is consistent with experimental evidence,  since it is not reported the emission of a dinucleons in heavy nuclei with spherical symmetry.
On the other hand, estimates of half life time show satisfactory results and replicate all the half life times for different even-even heavy nuclei with spherical symmetry, by adjusting the $\left\langle n\right\rangle$ as free parameters. 
Note that the model presented reproduces the half-life times, without resorting to traditional semi-classical model.
The model achieved to justify the presence of the photon observed in $\alpha$ decay processes also corroborates the fact that it has very low energy, as only taking it into account through angular coupling, reproduces the half life times of large numbers nuclei, but also provides an explanation of why there is not possibly in some processes.

\end{document}